\documentclass[10pt,conference]{IEEEtran}
 \IEEEoverridecommandlockouts
\usepackage[utf8]{inputenc}
\usepackage{cite}
\usepackage{svg}
\usepackage{graphicx}
\usepackage{epstopdf}  % Converts EPS to PDF
\DeclareGraphicsExtensions{.pdf,.png,.jpg,.eps}
\usepackage{booktabs}
\usepackage{listing}
\usepackage{xcolor}  % for colored text
\usepackage{tcolorbox}  % for the box
\usepackage{enumitem}  % for list formatting
\usepackage{fancybox}  % for fbox
\usepackage{array}  % for better table and minipage handling
\usepackage{listings}
\usepackage{comment}
\usepackage{hyperref}
\hypersetup{
    citecolor=black,
    colorlinks=true,
    linkcolor=blue,
    filecolor=blue,      
    urlcolor=blue,
    linkbordercolor={1 1 1}
    }
\author{
\IEEEauthorblockN{Madhurima Chakraborty\IEEEauthorrefmark{1}}
\IEEEauthorblockA{mchak009@ucr.edu \\
University of California, Riverside \\
CA, USA}
\thanks{\IEEEauthorrefmark{1} This work was performed under the auspices of the U.S. Department of Energy by Lawrence Livermore National Laboratory under Contract DE-AC52-07NA27344. LLNL-CONF-2002128}
\and
\IEEEauthorblockN{Peter Pirkelbauer, Qing Yi}
\IEEEauthorblockA{pirkelbauer2@llnl.gov, yi7@llnl.gov  \\
Lawrence Livermore National Laboratory (LLNL)\\
CA, USA}
}

\newcommand{\formal}[1]{\texttt{#1}} % One parameter
\tcbuselibrary{skins}
\title{\formal{FormalSpecCpp}: A Dataset of \formal{C++} Formal Specifications created using LLMs}
\begin{document}

\maketitle

\begin{abstract}
%This paper introduces FormalSpecCpp, a dataset to address the critical need for standardized verification of the C++ program specifications. To the best of our knowledge, this is the first comprehensive collection of C++ programs with well-defined formal specifications. Additionally, we present a specification translation approach that employs rule-based prompting strategies to handle critical aspects of the target programming language such as appropriate type selection, safety checks, and assertion mapping. This work provides a standardized dataset for evaluating specification inference tools and explores how Large Language Models (LLMs) can support automated specification generation across various programming languages. We envision this benchmark advancing research on program verification, automated testing, and specification inference for C++ programs and providing guidelines for using LLMs to generate specifications for other programming languages. The dataset is  publicly available at \url{https://github.com/MadhuNimmo/FormalSpecCpp}.

\formal{FormalSpecCpp} is a dataset designed to fill the gap in standardized benchmarks for verifying formal specifications in \formal{C++} programs. To the best of our knowledge, this is the first comprehensive collection of \formal{C++} programs with well-defined preconditions and postconditions. It provides a structured benchmark for evaluating specification inference tools and testing the accuracy of generated specifications.
Researchers and developers can use this dataset to benchmark specification inference tools, fine-tune Large Language Models (LLMs) for automated specification generation, and analyze the role of formal specifications in improving program verification and automated testing. By making this dataset publicly available, we aim to advance research in program verification, specification inference, and AI-assisted software development. The dataset and the code are available at \url{https://github.com/MadhuNimmo/FormalSpecCpp}.
\end{abstract}

%This paper introduces \formal{FormalSpecCpp}, a data set to address the critical need for standardized verification of the specifications of the \formal{C++} program. To our knowledge, this is the first comprehensive collection of \formal{C++} programs with well-defined formal specifications. Additionally, we present a specification translation approach that employs rule-based prompting strategies to handle critical aspects of the target programming language such as appropriate type selection, safety checks, and assertion mapping. This work provides a standardized dataset for evaluating specification inference tools and explores how Large Language Models (LLMs) can support automated specification generation across various programming languages. This benchmark is available at \href{https://madhunimmo.github.io/}{madhunimmo.github.io} for advancing research on program verification, automated testing, and specification inference for \formal{C++} programs and providing guidelines for using LLMs to generate specifications for other programming languages.

\begin{figure*}[htp!]
    \centering    \includegraphics[width=0.85\textwidth,keepaspectratio]{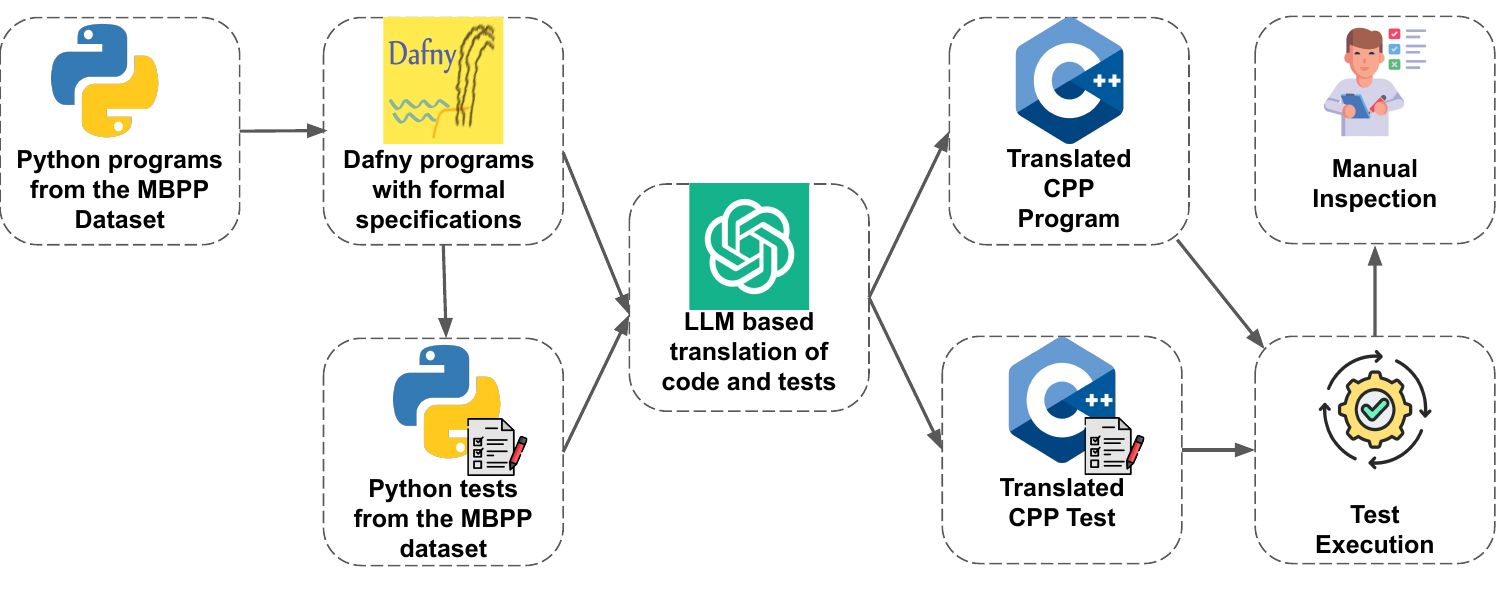}
    \caption{Pipeline for translating \formal{Python} programs and tests from the MBPP dataset to \formal{C++} using \formal{Dafny} formal specifications and LLM-based translation.}
    \label{fig:workflow}
\end{figure*}

\section{Introduction}
Preconditions and postconditions serve as formal specifications that define the expected state of a program before and after code execution. This ensures software correctness, formal verification, and code reliability \cite{meyer1992design,hoare1969axiomatic}. Despite their significance, standardized benchmarks for \formal{C++} programs with formal specifications have been notably absent due to the lack of standardization, the complexity of \formal{C++}, the evolving language standards, and the limited adoption of formal contracts. There is a growing need for formal specification support within programming languages to improve software reliability. Historically, \formal{C++} lacked support for formal specifications. Recognizing their importance, the \formal{C++} Standards Committee committed to introducing formal specifications in \formal{C++26}~\cite{doumler2024contracts}. This will standardize the ability to specify function contracts directly in the language.
For example, refer to the code below:

%\vspace{-0.5em}
{\fontsize{7.5pt}{8pt}
\begin{verbatim}
int f(const int x)
    pre (x != 1) // a precondition assertion
    post(r : r != 2) // a postcondition assertion
{
    contract_assert(x != 3); // an assertion statement
    return x;
}
\end{verbatim}
}
%\vspace{-1em}
This feature represents a better alternative to the current practice of using macros or comments for specifying contracts, allowing developers to express expectations more clearly and formally. It also enhances the reliability of the code, facilitates better documentation of function behavior, and supports run-time behaviors for contract verification. Although this is promising for newer codebases, manually adapting it to the scale of legacy \formal{C++} is prohibitively expensive.

Using LLMs to generate formal specifications is a viable solution to this bottleneck. Given that such specifications can be verified against existing test suites, they enhance code reliability without requiring extensive manual effort. This work highlights the potential of LLMs in supporting automated specification generation, addressing the need for tools that can infer formal specifications from code. Our dataset enables an empirical evaluation of the specification inference tools. By comparing inferred specifications against ground truth, researchers can assess precision, recall, and correctness in automated inference methods.

%The FormalSpecCpp dataset is derived from the \formal{Dafny}-synthesis benchmark~\cite{misu2024towards}, which provides formally verified \formal{Dafny} programs with preconditions and postconditions. Using prompt-driven translation, we systematically convert these \formal{Dafny} programs into \formal{C++} implementations while preserving their formal specifications. The prompting-based approach employs the OpenAI GPT-4-turbo LLM for systematic translation of code along with formal specifications from \formal{Dafny} to \formal{C++}. 

The \formal{FormalSpecCpp} dataset is derived from the \formal{Dafny}-synthesis benchmark~\cite{misu2024towards}, which provides formally verified \formal{Dafny} programs with preconditions and postconditions. We use a prompt-driven approach with OpenAI's \textbf{GPT-4-turbo} model to systematically translate these \formal{Dafny} programs into \formal{C++}, ensuring that the generated code preserves its formal specifications.

% The significance of this work goes beyond translation. By providing a standardized set of \formal{C++} programs with verified specifications, we enable
% \textbf{(a)} evaluation and comparison of specification inference tools using a consistent benchmark, and
% \textbf{(b)} exploration of how LLMs can support automated specification generation across diverse programming languages.

This dataset is the ground truth benchmark for evaluating formal specification inference tools by providing verified \formal{C++} programs with preconditions and postconditions. It also enables the testing of formal verification tools to assess how well they enforce these specifications. Additionally, it supports static and dynamic analysis research, helping researchers examine the impact of different specification techniques on software correctness, tool performance, and contract validation in real-world \formal{C++} programs.

Our prompt engineering methodology systematically translates \formal{Dafny} specifications into \formal{C++} by explicitly handling type mapping, assertion transformation, and safety constraints. This structured approach reduces hallucinations, improves syntactic correctness, and ensures that the translated contracts accurately represent program intent.

\section{Background}

This section briefly overviews the background details: formal specifications in software engineering, LLMs and prompt engineering.

% \subsection{Formal Specifications}  

% Formal specifications are precise mathematical descriptions of software behavior, defining expected states and outcomes of a program at various points of execution. They play a critical role in ensuring software correctness, enhancing maintainability, and supporting formal verification processes~\cite{meyer1992design, hoare1969axiomatic}. By explicitly stating pre-conditions (what must hold true before execution) and post-conditions (what will be true after execution), these specifications establish a clear contract between different components of a program.  

% Formal specifications provide several key benefits:  
% \begin{itemize}
%     \item Enhanced Code Reliability: By defining exact requirements and guarantees, they help developers write safer and more predictable code.  
%     \item Modular Reasoning: These specifications enable developers to reason about individual code units without diving into their internal implementations.  
%     \item Facilitation of Formal Verification: They allow automated tools to verify program properties, catching issues during development rather than in production. 
% \end{itemize}  

% However, writing formal specifications is often challenging, particularly for large and legacy codebases. This has driven research into automating the inference of these specifications, addressing the cost and time overhead associated with manual specification writing.  

\subsection{Formal Specifications}  

Formal specifications define the expected behavior of software using precise mathematical statements, typically through preconditions (requirements before execution) and postconditions (guarantees after execution). They play a critical role in software correctness, modular reasoning, and formal verification~\cite{meyer1992design, hoare1969axiomatic}.  

Despite their benefits, manual specification writing is complex and time-consuming, especially for large or legacy codebases. This has led to increasing research in automating specification inference, reducing the cost of manual annotations while improving software reliability.  

\subsection{Large Language Models (LLMs)}
LLMs are transformer-based neural networks trained on vast text corpora, enabling them to generate human-like text and code. These models, such as GPT-4, excel in code synthesis, reasoning, and text generation, making them valuable tools for automating formal specification inference.  

LLMs can be broadly categorized into:
\begin{itemize}
    \item \textbf{General-purpose models} (e.g., GPT-4, Claude) trained on diverse text and code.
    \item \textbf{Code-specific models} (e.g., Code Llama) optimized for programming tasks.
\end{itemize}

For our dataset development, we selected GPT-4-turbo, a variant of OpenAI’s GPT models, known for its strong performance in generating semantically correct code and a large 128k-token context window~\cite{openai_gpt4turbo_docs}. Prompt engineering is used to refine the model inputs to systematically guide the synthesis of formal specifications.

% Large Language Models (LLMs) are a class of artificial intelligence models designed to process and generate human-like text. These models, such as GPT-4, are built upon transformer architectures, leveraging billions of parameters trained on diverse text datasets. LLMs excel in tasks such as code synthesis, language translation, and text summarization, owing to their ability to learn contextual representations of language.  These models can be broadly categorized into:
% \begin{itemize}
% \item General-purpose models (e.g., GPT-4, Claude) that excel at understanding both natural language and code
% \item Code-specific models (e.g., Code Llama) optimized for programming tasks
% \end{itemize}
% For our dataset development, we selected GPT-4-turbo, a variant of the GPT (Generative Pre-trained Transformer) family of models developed by OpenAI. The GPT models have undergone rapid evolution, from GPT-2 to the latest iterations like GPT-4, bringing significant improvements in language understanding, contextual reasoning, and generation accuracy. In our work, GPT-4-turbo was chosen for its exceptional performance in generating both functionally and semantically correct code and a substantially large  context window of 128k tokens \cite{openai_gpt4turbo_docs}. Through prompt engineering, we systematically crafted and refined input prompts to leverage GPT-4's capabilities for automating formal specification generation, demonstrating its potential in this domain.

\subsection{Prompt Engineering}
Prompt engineering involved designing structured inputs to guide LLMs toward accurate and relevant outputs. Effective prompts improve clarity, control structure, and mitigate issues like hallucinations or incorrect type inferences.  

In our work, prompt engineering plays three critical roles:
\begin{itemize}
    \item \textbf{Specification Preservation}: Ensuring formal properties are accurately translated between languages.
    \item \textbf{Type Safety}: Mapping \formal{Dafny} types to appropriate \formal{C++} equivalents.
    \item \textbf{Assertion Mapping}: Converting \formal{Dafny} verification constructs into valid \formal{C++} assertions and macros.
\end{itemize}

By carefully designing prompts, we systematically transformed \formal{Dafny} programs with verified specifications into \formal{C++} implementations while preserving formal correctness.

\begin{figure*}[h]
    \centering
    \includegraphics[width=0.85\textwidth]{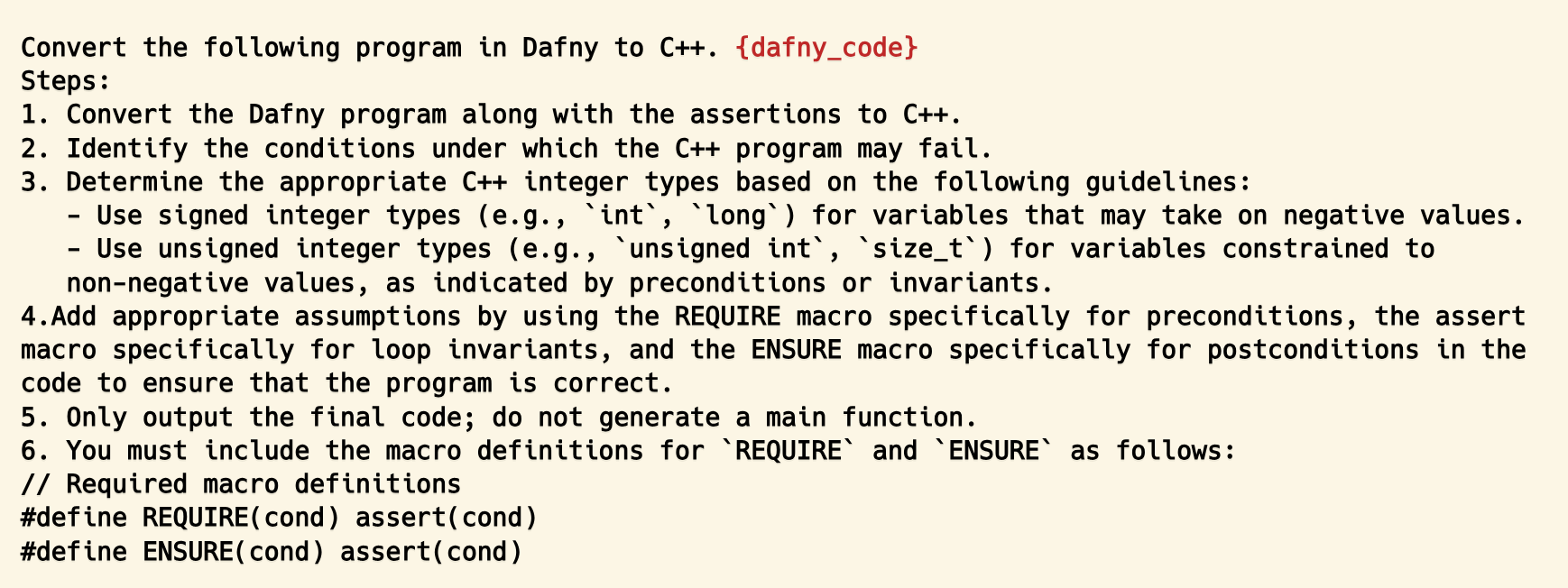}
    \caption{\formal{Dafny} to \formal{C++} Code Translation Prompt}
    \label{fig:prompt}
\end{figure*}

\section{The \formal{FormalSpecCpp} Dataset}

This section describes the steps to generate the \formal{FormalSpecCpp} dataset. Figure \ref{fig:workflow} presents the workflow.

\subsection{Conversion of \formal{Dafny} Programs to \formal{C++}}
% The benchmark \formal{Dafny}-synthesis~\cite{misu2024towards}, which contains verified \formal{Dafny} programs with specifications, serves as the starting point for generating \formal{C++} programs. Their study intended to evaluate how well LLMs can synthesize the \formal{Dafny} code, including formal specifications and validation conditions that pass the \formal{Dafny} verifier. Our goal is to create a benchmark of \formal{C++} programs with formally verified specifications. While their approach relies on few-shot prompting and the LLM itself to generate specifications, we hypothesized that a more reliable and straightforward method is to translate verified \formal{Dafny} programs including their formal specifications directly into \formal{C++}. This approach reduces the dependence on the LLM for generating specifications, leveraging its strengths for translation instead. By using verified programs with preconditions and postconditions, we ensure that formal correctness is preserved, creating a consistent and reliable benchmark. This minimizes potential inaccuracies introduced by the generative process, resulting in a more deterministic and reliable benchmarking pipeline. Since \formal{C++} does not natively support preconditions and postconditions yet, we defined a custom format to represent them consistently across all programs:

The \formal{Dafny}-synthesis benchmark~\cite{misu2024towards}, which contains verified \formal{Dafny} programs with formal specifications, is the starting point for generating \formal{C++} programs. Their study evaluated how well LLMs can synthesize \formal{Dafny} code, including formal specifications and validation conditions that pass the \formal{Dafny} verifier.

% While their approach relies on few-shot prompting with the LLM to generate formal specifications, we opted for a different strategy. Instead of inferring formal specifications directly for \formal{C++} using LLMs, we used LLMs for a structured translation process from \formal{Dafny} to \formal{C++}. This decision was based on the following key considerations:
% \begin{itemize}
%     \item \textbf{Formal correctness preservation}: Since \formal{Dafny} programs in the benchmark are already verified, translating their specifications directly to \formal{C++} ensures that formal and logical correctness is maintained rather than  relying on speculative inference by the LLM.
%     \item \textbf{Reduced LLM hallucinations}: <<change>>Direct specification inference for \formal{C++} would require the LLM to generate preconditions and postconditions based on assumptions about the code’s intended behavior. This increases the risk of incorrect or inconsistent specifications, as inferred contracts may not align with the actual program semantics.
% \end{itemize}
While their approach relies on few-shot prompting with the LLM to generate formal specifications \cite{misu2024towards}, we opted for a different strategy. Instead of inferring formal specifications directly for \formal{C++} using LLMs, we used LLMs for a structured translation process from \formal{Dafny} to \formal{C++}. This decision was based on the following key considerations:

\begin{itemize}
    \item \textbf{Formal correctness preservation}: Since \formal{Dafny} programs in the benchmark are already verified, translating their specifications directly to \formal{C++} ensures that formal and logical correctness is maintained rather than relying on speculative inference by the LLM.
    
    \item \textbf{Avoiding prompt dependence and unnecessary constraints}: Prior work \cite{misu2024towards} found that effective specification inference with LLMs requires carefully engineered prompts, yet the results are often redundant or overly restrictive. By translating verified \formal{Dafny} specifications, we eliminate reliance on complex prompt tuning and prevent the introduction of unnecessary constraints.
\end{itemize}

By structuring the translation process around an already verified source, we minimize potential inaccuracies introduced by the generative process, resulting in a more deterministic and reliable benchmarking pipeline. We considered only those programs from \formal{Dafny}-synthesis that had both preconditions and postconditions defined, totaling 105 programs. Since \formal{C++} does not yet natively support preconditions and postconditions, we defined a custom format to represent them consistently across all programs:
{
\fontsize{7.5pt}{8pt}
\begin{verbatim}
#define REQUIRE(cond) assert(cond)
#define ENSURE(cond) assert(cond)
\end{verbatim}
}
The final prompt for converting the programs is shown in Figure \ref{fig:prompt}.

\subsection{Conversion of Associated Tests}
The \formal{Dafny}-synthesis benchmark includes test cases that were manually translated from \formal{Python} tests in the MBPP dataset~\cite{austin2021program}. In contrast, we used LLM to directly convert the original MBPP test cases into \formal{C++}, translating test input, expected output, and assertions while maintaining functional equivalence.

\subsection{Automated Testing of Translated Programs}

After generating the \formal{C++} programs and their corresponding tests, we performed automated test execution to validate their correctness. This step revealed several types of errors, including:  
\textbf{(a)} inconsistencies in preconditions and postconditions,  
\textbf{(b)} incomplete or ill-formed test cases that failed to capture intended functionality, and  
\textbf{(c)} mismatches between expected and actual outputs, indicating discrepancies in program behavior.  

The automated tests provided valuable feedback, identifying functional issues in the translated programs. In our initial run, \texttt{62} programs passed all tests, while \texttt{43} failed due to assertion errors, linker errors, and compilation errors. Further analysis showed that many assertion failures stemmed from incorrect test specifications, such as misordered arguments, incorrect expected values, or type mismatches (\formal{unsigned int} vs. \formal{double}). Compilation errors, on the other hand, often reflected the limitations of LLM-based translation of the tests.  

Although automated tests identified structural defects, they could not ensure semantic correctness. Some programs were compiled successfully but contained incorrect specifications or logical inconsistencies that required manual review.  

\subsection{Manual Verification of Translations}  
To address these shortcomings, one of the authors, an expert in formal specifications and \formal{C++} programming, manually reviewed and refined a subset of translated programs and test cases. This process focused on:  
\begin{itemize}  
    \item Correcting errors detected by automated tests, such as misaligned assertions, type mismatches, and faulty preconditions.  
    \item Inspecting a subset of successful programs to verify semantic correctness and adherence to best practices.  
    \item Refining specifications and tests to better capture intended function behavior.  
\end{itemize}  

Manual verification uncovered errors that automated methods missed, particularly in cases where specifications were syntactically valid but logically incorrect. These included logical inconsistencies such as misaligned assertions and faulty precondition constraints. This step improved the dataset’s reliability by ensuring that specifications accurately reflected program behavior.

The \formal{FormalSpecCpp} dataset is available at \href{https://github.com/MadhuNimmo/FormalSpecCpp}{https://github.com/MadhuNimmo/FormalSpecCpp}. It serves as a standardized dataset for evaluating and improving precondition and postcondition inference in \formal{C++}. It will be useful in advancing research in specification verification and automated program analysis.

\section{Cost Analysis and Efficiency}  

The conversion process from \formal{Dafny} to \formal{C++} programs and tests was highly efficient, as summarized in Table \ref{tab:overall_statistics}. A total of 105 files with preconditions and postconditions were processed, with all converted programs successfully compiling on the first attempt, achieving a $100\%$ success rate. The entire process was completed in approximately 27 minutes, demonstrating the scalability and speed of the approach.

%The conversion process from \formal{Dafny} to \formal{C++} programs and tests was highly efficient, as summarized in Table \ref{tab:overall_statistics}. 105 files with preconditions and postconditions were processed with a success rate of $100\%$, achieving all conversions on the first attempt. The entire process was completed in approximately 27 minutes, demonstrating the scalability and speed of the approach.  
The total cost for converting these files was \formal{\$2.07}, averaging \formal{\$0.02} per file. Similarly, generating the associated test cases cost an additional \formal{\$1.31} and required around 15 minutes. These results underscore the minimal resource usage and cost-efficiency of the pipeline.  
\formal{Dafny} files averaged \formal{17.3} lines of code, while the converted \formal{C++} code showed a $91.2\%$ increase in size, with an average of \formal{33.1} lines per file.

This workflow's smooth transition and minimal resource overhead demonstrate its potential for practical use in large-scale datasets and real-world applications.

    \begin{table}[h]
        \centering
        \caption{Overall Statistics}
        \label{tab:overall_statistics}
        \begin{tabular}{@{}ll@{}}
            \toprule
            \textbf{Statistic}                          & \textbf{Value}          \\ \midrule
            Total files processed                       & 105                     \\
           Successfully compiled programs(1st attempt)                   & 105 (100\%)             \\
            Total cost                                  & \$2.07                  \\
            Average cost per file                       & \$0.02                  \\
            Average \formal{Dafny} lines                         & 17.3                    \\
            Average \formal{C++} lines                           & 33.1                    \\
        \bottomrule
        \end{tabular}
    \end{table}

\section{Related Works}

\subsection{Formal Methods and Specifications}
Hoare's axiomatic basis for computer programming~\cite{hoare1969axiomatic} established the foundation for reasoning about program correctness through preconditions and postconditions, while Dijkstra's weakest precondition calculus~\cite{dijkstra1976discipline} provided systematic approaches for deriving specifications. These theoretical frameworks influenced the development of specification languages like JML~\cite{leavens2008jml} and modern verification-oriented programming languages such as \formal{Dafny}~\cite{leino2010Dafny}. Meyer's design by contract~\cite{meyer1992design} further popularized the use of preconditions and postconditions in software development. However, the manual effort required to write formal specifications has historically limited their widespread adoption in industrial practice~\cite{woodcock2009formal}.

\subsection{Automated Specification Inference}
The challenge of manual specification writing has motivated extensive research in automated specification inference. Early approaches like Daikon pioneered dynamic analysis techniques to infer likely program invariants\cite{ernst2007daikon}.
Static analysis methods, including abstract interpretation~\cite{cousot1977abstract} and symbolic execution~\cite{king1976symbolic}, provided more comprehensive approaches but faced scalability challenges with complex codebases~\cite{calcagno2011compositional}.

Recent research has demonstrated the potential of LLMs in generating specifications and code for formal verification languages like \formal{Dafny} \cite{misu2024towards} and Java \cite{ma2024specgen}. This trend aligns with broader efforts to leverage AI in software engineering tasks, particularly in areas requiring formal reasoning.

%Misu et al. \cite{misu2024towards} investigated the capabilities of LLMs in synthesizing verified \formal{Dafny} methods, showing that with appropriate prompting techniques, models like GPT-4 could generate correct and verifiable \formal{Dafny} code for a significant portion of test problems. Similarly, Ma et al. \cite{ma2024specgen} explored the use of LLMs for generating formal specifications in Java, demonstrating the models' ability to produce meaningful specifications for Java methods.

%The conversational prompt style, inspired by works like Xia et al. \cite{xia2024automated} has proven effective in eliciting more detailed and context-aware responses from LLMs. This approach allows for a more natural interaction with the model, potentially leading to more accurate and comprehensive code generation.

%The ability of GPT-4-turbo to generate executable code in a single attempt for all considered programs represents a significant advancement in LLM capabilities for code synthesis. This efficiency aligns with the growing body of research demonstrating the potential of LLMs in various software engineering tasks.

%The success in test case generation by LLMs, as shown by Siddiq et al. and Alshahwan et al. \cite{siddiq2024using,alshahwan2024automated}, further underscores the potential of these models in formal methods and software testing. These studies highlight how LLMs can be leveraged to automate and enhance critical aspects of the software development lifecycle, from specification writing to test case generation.

\section{Concluding Remarks}
This paper introduces the \formal{FormalSpecCpp} dataset with formal specifications for \formal{C++}, encouraging the community to utilize it for diverse research applications. Although the benchmark is a significant step forward, there is room for improvement. We highlight some of the challenges and lessons learned along the way. This is meant to serve as a guidance for researchers interested in extending the approach.

\subsection{Code Translation}
The prompt to convert \formal{Dafny} programs into valid and semantically correct \formal{C++} code required non-trivial iterative refinement, especially given the differences in their type systems. For instance, \formal{Dafny} supports an unbounded integer type, which can represent any integer value without overflow, whereas \formal{C++} offers multiple fixed-size integer types, including \formal{int}, \formal{unsigned int}, \formal{short}, and \formal{long}. \formal{C++} has a well-defined overflow behavior for signed integers and wraps around for unsigned integers. One should be mindful of these changes when translating from one programming language to another.

%The prompt to convert \formal{Dafny} programs into valid and semantically correct \formal{C++} code required a non-trivial iterative refinement, especially given the differences in their type system. For instance, \formal{\formal{Dafny}} supports unbounded integer type which can represent any integer values without overflow, but \formal{C++} offers multiple fixed-size integer types including \formal{int}, \formal{unsigned int}, \formal{short}, and \formal{long}. \formal{C++} has a well-defined overflow behavior for signed integers and wraps around for unsigned integers. One should be mindful of these changes When translating from one programming language to another. 

\subsection{Prompt Engineering Refinements}  
Crafting effective prompts for \formal{\formal{Dafny}} to \formal{C++} translation involved refining the language model's handling of preconditions, postconditions, and correct \formal{C++} constructs. Early iterations resulted in misaligned assertions and syntax errors due to differences in language semantics. The iterative process focused on explicitly specifying rules around types, preconditions, invariants, and postconditions, ensuring consistency and correctness across the translations.

\subsection{Test Case Translation}
Adapting test cases from \formal{\formal{Dafny}} to \formal{C++} required a structured approach to facilitate compatibility with \formal{C++} testing frameworks. We designed prompts to guide the LLM in generating accurate \formal{C++} test cases in a predefined format, reducing manual effort and ensuring smooth integration into the \formal{C++} testing environment.

Refining these prompts provided valuable insights into the strengths and limitations of using LLMs for test case translation. While improved prompts enhanced the LLM’s handling of complex cases, challenges remained, such as occasional misinterpretation of logical relationships or the generation of syntactically correct but semantically flawed code.

\subsection{Future work} Some natural extensions to this work are
(a) including additional programs and programming languages,
(b) refining the prompt engineering process,
(c) exploring reinforcement learning and hybrid approaches to further enhance the specification generation and validation, and
(d) extending the dataset to include larger programs with more complex specifications to test the scalability and robustness of the inference and validation processes.

% By addressing these aspects, we aim to advance research in verified program synthesis and formal specification generation.

%  \section*{Acknowledgments}
% This work was performed under the auspices of the U.S. Department of Energy by Lawrence Livermore National Laboratory under Contract \texttt{DE-AC52-07NA27344}. The United States Government retains, and the publisher, by accepting the article for publication, acknowledges that the United States Government retains a non-exclusive, paid-up, irrevocable, world-wide license to publish or reproduce the published form of this manuscript, or allow others to do so, for United States Government purposes.

\bibliographystyle{IEEEtran}
\bibliography{references}

\end{document}